# Superconductivity at 5.2 K in ZrTe$_3$ polycrystals and the effect of Cu, Ag intercalation

C. S. Yadav and P. L. Paulose
*Department of Condensed Matter Physics and Materials Science,
Tata Institute of Fundamental Research, Colaba, Mumbai-400005, India*

We report the occurrence of superconductivity in polycrystalline samples of ZrTe3 at 5.2 K temperature at ambient pressure. The superconducting state coexists with the charge density wave (CDW) phase, which sets in at 63K. The intercalation of Cu or Ag, does not have any bearing on the superconducting transition temperature but suppresses the CDW state. The feature of CDW anomaly in these compounds is clearly seen in the DC magnetization data. Resistivity data is analysed to estimate the relative loss of carriers and reduction in the nested Fermi surface area upon CDW formation in the ZrTe$_3$ and the intercalated compounds.

Interplay of Charge Density Wave and superconducting (SC) states continues to be a subject of significant interest. [1] Low dimensional chalcogenides of transition metals are one of the well studied systems that show the coexistence of these competing phenomena. 2H-NbSe$_2$ shows the CDW at 32 K and SC at 7 K, whereas Cu intercalation of 1T-TiSe$_2$ suppresses CDW and introduces SC at low temperature.[2, 3] In the recent reports Copper and Nickel intercalated compounds of ZrTe$_3$ are shown to exhibit bulk superconducting properties at 3.8 K and 3.1 K respectively.[4, 5] The parent compound ZrTe$_3$ shows SC transition below 2 K but the volume of SC fraction is very small (< 5 %).[6] It has unique crystal structure among the IV-trichalcogenides and has triangular prismatic chains. There are two identical chains connected by inversion symmetry in the monoclinic unit cell, in such a way that neighboring chains (parallel to the b-direction), make alternate chain pairs.[7, 8, 9] The electrical resistivity of ZrTe$_3$ single crystals shows metallic behavior with anisotropy ratio of 1:1:10 for $\rho_a$, $\rho_b$, $\rho_c$ respectively.[7] The metallic properties in ZrTe$_3$ emerges as a result of enhanced overlap between bands of the different chains (interchain Zr-Te distance is shorter than the intrachain distance). The lesser difference in the electro-negativities of Zr and Te compared to other chalcogen atoms, makes more electrons available for the conduction.[6, 10]

The electrical resistivity of ZrTe$_3$ measured along the different directions show the CDW anomaly at $T_{CDW}$ = 63 K in 'a' and 'c' directions but not in 'b' direction.[6, 7] Photoemission studies and band structure calculations attribute this behavior due to the nesting in a small electron pocket of the highly directional Te-Te chains while other sheets of the Fermi surface remain unaffected. [11] The opening of the gap in the electronic dispersion of the band follow the BCS model of Peierls transition with a mean field transition temperature about four times higher than $T_{CDW}$. [12] Dependence of CDW transition on pressure has been very unusual. $T_{CDW}$ initially increases to 114 K at 1 GPa and then decreases monotonously before abruptly vanishing at 5 GPa. [13, 14] The increase in the pressure enhances three dimensionality of the structure and thus diminishing the area of the planar portion of the Fermi surface (FS) resulting in the reduction of the nested FS. The effect of pressure on CDW is akin to that of the doping of the foreign atoms in the lattice, where redistribution of charge leads to the reduction in the nested FS. [13, 14]

Previous studies have reported the presence of filamentary SC in single crystalline ZrTe$_3$ with $T_c$ < 2 K, with weakly coupled SC filaments aligned parallel to the 'a' axis with the spacing of 4 nm. [6, 7] These SC filaments run parallel to the 'a' axis, perpendicular direction of triangular prismatic chain of ZrTe$_3$. There is strong pressure dependence on $T_c$ as in the case of $T_{CDW}$. $T_c$ initially falls below 1.2 K at 0.5 GPa, then increases again to 4.7 K at 11 GPa. [13] Since the coupling constant for SC pairing is proportional to the electronic constant divided by the mean square phonon energy, the softening of the phonon modes enhances the coupling constant and increase the $T_c$. [14, 15]

In the literature, most of work is focused on single crystalline samples and there is no report on the polycrystalline ZrTe$_3$, highlighting the superconducting or charge density wave properties. Recent report has shown bulk superconductivity in the Cu and Ni intercalated ZrTe$_3$ crystals with the $T_c$ of 3.8 K and 3.1 K respectively. [4, 5] In the present work we have undertaken the study on polycrystalline ZrTe$_3$, and intercalation with small amount of Cu and Ag. The polycrystal samples have one advantage of homogeneous distribution of intercalants compared to the crystals grown by vapor transport method. It becomes even more important when the intercalant percentage is very low as in our case (5 %). The main finding of the present work is the occurance of superconductivity at enhanced temperature of 5.2 K in polycrystallline ZrTe$_3$ in ambient pressure. Though superconductivity is still filamentary, just as for the single crystals. Unlike for the single crystals, the Cu or Ag intercalations are not able to enhance the $T_c$ in polycrystal samples. We have also studied the effect of intercalant on the CDW state of the compound. The CDW anomaly is clearly visible in the high field magnetization data in these weakly magnetic materials. The charge transfer by the intercalant Cu and Ag, is found to enhance the electrical conductivity of the ZrTe3 and suppress the CDW anomaly.

Polycrystalline samples of ZrTe$_3$, Cu$_{0.05}$ ZrTe$_3$, and Ag$_{0.05}$ ZrTe$_3$} were prepared using high purity elements Zr (99.9%), Te (99.9%), Cu (99.999%), and Ag (99.9%) taken in the stoichiometric ratio, inside evacuated ($10^{-6}$ mbar pressure) quartz tubes at 975$^0$C for 48 hours. The reacted material contained micron size small crystallites. For resistivity and heat capacity measurements, the compounds were pelletized at 10 ton pressure. The obtained pellets were of density ~ 6 gm/cm$^3$. The electrical contacts were made using the high quality silver paint and contact resistances were found to be 1-2Ω. The X Ray-diffraction patterns of the compounds were taken by Philips X'pert PRO Diffractometer in Brag Brentano geometry. Quantum Design- SQUID Magnetometer was used for magnetic measurement and QD-PPMS (Physical Properties Measurement System) was used for resistivity and heat capacity measurements.

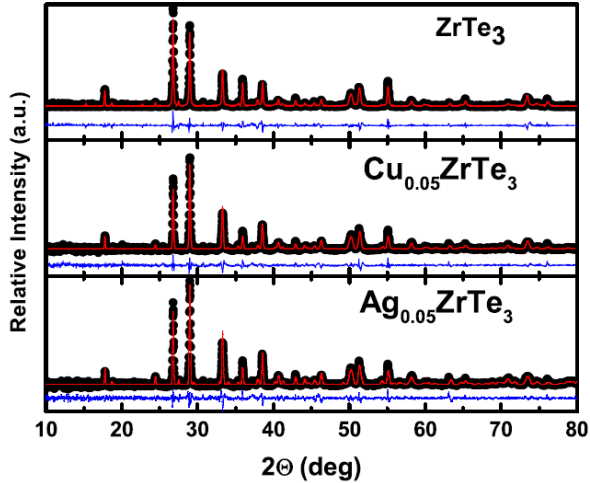

Fig.1 (Color online) The rietveld refined X Ray diffraction patterns for the ZrTe$_3$, Cu$_{0.05}$ ZrTe$_3$, and Ag$_{0.05}$ZrTe$_3$. Black circles show observed pattern, Red line shows calculated pattern and the Blue line represents the difference between observed and calculated patterns.

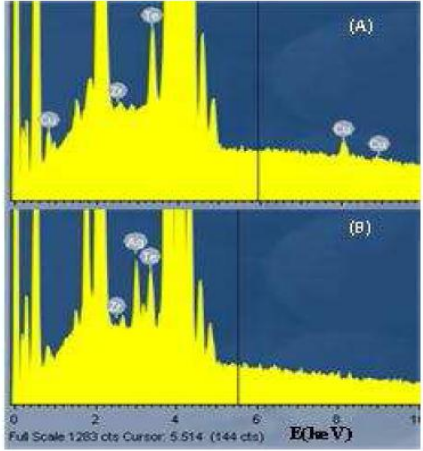

Fig.2 (Color online) SEM-EDS profiles for the Cu$_{0.05}$ZrTe$_3$ (upper), and Ag$_{0.05}$ZrTe$_3$ (lower) compounds showing the peak intensities for the Cu, Ag, Zr and Te atoms. Un-marked and high intensity peaks are from zirconium and tellurium atoms only.

The X Ray diffraction pattern of the ZrTe$_3$, Cu and Ag intercalated ZrTe$_3$ compounds were analyzed using Rietveld (GSAS) program and was found to fit with the monoclinic (space group P2$_1$/m) structure (Figure 1). The lattice parameters as obtained from the Rietveld fit are a = 5.863Å, b = 3.923Å, c = 10.089Å, and β = 97.74$^0$ for ZrTe$_3$. It is difficult to detect any appreciable change in the lattice parameters for the 5% intercalation of Cu/Ag in ZrTe$_3$ using our X ray diffraction data. It is unlikely that the intercalant atoms are substituting the Zr atom because of the big mismatch in the atomic size. However the change in the relative peak intensities is one signature of intercalation in the compound. A good quality neutron diffraction pattern is required to see the effect of small amount intercalants on the lattice parameters. The samples were characterized using SEM-EDX (Scanning Electron Microscope- Energy Dispersive X-Ray spectroscopy), to ascertain the composition of the compound, which were found to be same as desired stoichiometry, within the instrument accuracy limit (~ 2%). Figure 2 shows the EDX graph for Cu$_{0.05}$ZrTe$_3$, and Ag$_{0.05}$ZrTe$_3$, clearly showing the presence of Cu and Ag atoms in the compounds.

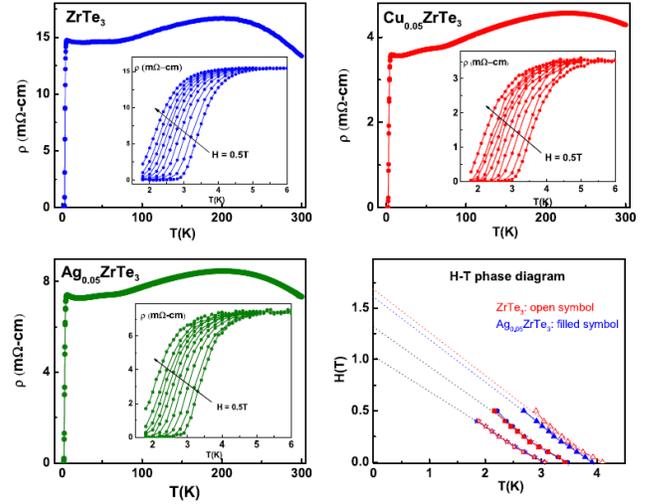

Fig. 3 (Color online) Temperature dependence of the electrical resistivity of ZrTe$_3$ (a), Cu$_{0.05}$ZrTe$_3$ (b), and Ag$_{0.05}$ZrTe$_3$ (c); showing the SC T$_c$. The field dependence of T$_c$ is shown in the inset of the respective figures for H = 0, 0.05, 0.10, 0.15, 0.20, 0.25, 0.30, 0.35, 0.40, and 0.50T. The HT phase diagram for ZrTe$_3$ and Ag$_{0.05}$ ZrTe$_3$ is shown in figure 2d. The dotted lines is the extrapolation of the linear H-T behavior to T = 0 K. The H$_{c2}$(0) values obtained are 1.6 T (ZrTe$_3$), 1.7 T (Ag$_{0.05}$ZrTe$_3$); 1.3 T ; and 1.0 T for the field values taken from 90%, 50%, and 10% of the ρ(T$_{onset}$).

The electrical resistivity of the ZrTe$_3$, Cu$_{0.05}$ZrTe$_3$, and Ag$_{0.05}$ZrTe$_3$ (figure 3), shows the onset of the SC transition (T$_c$) at 5.2 K. The higher value of T$_c$ for polycrystalline ZrTe$_3$, in comparison to < 2K for single crystals of previous studies, is highly remarkable. However transition width (ΔT$_c$) of these polycrystalline samples is large with ΔT$_c$ ~ 2K, and shows the large effect of grain boundary interaction on transition temperature. The normal state resistivity of these compounds initially shows the semiconducting behavior and becomes metallic below 200K. Similar behavior has observed in recently discovered Iron chalcogenide superconductors FeTe$_{1-x}$Se$_x$, where the weak localization of the charge carriers leads to the rise in the resistivity upon cooling. [16, 17, 18] The grain boundary effects in the polycrystalline samples become more important and affect the electrical properties. The room temperature resistivity of the compounds decreases upon intercalation of Ag and Cu. The room temperature resistivity values are 13 mΩ-cm for $ZrTe_3$, 7 mΩ-cm for $Ag_{0.05}ZrTe_3$, and 4.2 mΩ-cm for $Cu_{0.05}ZrTe_3$. The $ZrTe_3$ is highly anisotropic in nature and has high resistivity along along 'c' direction. The room temperature resistivity of polycrystalline $ZrTe_3$ sample is ~ 10 times higher than for the single crystalline of $ZrTe_3$ along 'c' axis. [6, 10] The polycrystallinity of samples does not have any bearing on the CDW phase of the compound, and $T_{CDW}$ for $ZrTe_3$ is found to be 63 K, same as for the single crystals. The further analysis of CDW state is presented in the later section.

The magnetic field dependence of the $T_c$ is shown in the inset of the respective figures. All the compounds show strong field dependence on the SC transition and have almost similar upper critical field ($H_{c2}$). The HT phase diagram for $ZrTe_3$ and $Ag_{0.05}ZrTe_3$ for the field values taken from the point where resistivity drops to 10%, 50% and 90% of the onset $T_c$ is shown in the figure 2d. The HT phase diagram for polycrystalline $Cu_{0.05}ZrTe_3$ sample is similar to that of $ZrTe_3$. The Werthamer-Helfand-Hohenberg (WHH) relation $\mu_0 H_{c2}(0) = -0.693T_c \cdot (dH_{c2}/dT)_{T_c}$ gave the values of $H_{c2}(0)$ = 1.3 T for $ZrTe_3$, $Cu_{0.05}ZrTe_3$, and $Ag_{0.05}ZrTe_3$ ($T_c$ is taken at the midpoint of transition), which is slightly higher than 1.2 T for $Ni_{0.05}ZrTe_3$ in H//a direction.[5] Though the HT diagram follow the linear behavior rather than the empirical relation $H_{c2}(T) = H(0)(1 - T/T_c)^2$, the value of $\mu_0 H_{c2}/k_B T_c$ = 0.23 T/K$^{-1}$, is within the Pauli's weak coupling limit of $\mu_0 H_{c2}/k_B T_c$ = 1.84 TK$^{-1}$ for the singlet pairing. The coherence length ($\xi$) using the Ginzburg-Landau (GL) formula $\xi = (\Phi_0/2\pi \mu_0 H_{c2})^{1/2}$, where $\Phi_0 = 2.07 \times 10^{-7}$ Oe.cm$^2$ is calculated as 14.8 nm, which is comparable to that observed for the Cu and Ni intercalated $ZrTe_3$ single crystals.[4, 5]

The low field (H = 20 Oe) DC magnetization (ZFC and FC) data of the $ZrTe_3$, $Cu_{0.05}ZrTe_3$, and $Ag_{0.05} ZrTe_3$ is shown in the figure 4. The SC transition is broad and the onset $T_c$ is lower ($T_c$= 3.8 K) than observed in the electrical resistivity data. The estimated superconducting volume fraction for these compound is as low as 6%, 5%, and 8% for $ZrTe_3$, $Cu_{0.05}ZrTe_3$, and $Ag_{0.05}ZrTe_3$ respectively. Smaller SC volume fraction and broad transition temperature are suggestive of the filamentary nature of superconductivity and the distribution of $T_c$ within the material. We think that like the CDW state, which is observed in 'a' and 'c' directions only, superconductivity may also sets in only along these directions in the remnant part of the Fermi surface, and therefore is not a bulk feature. The MH hysteresis loop (not shown here) at 1.8 K is like the typical butterfly loop of type II superconductors. The MH loop at 1.8K for all the systems showed large jump in M upon field reversal which is indicative of higher value of critical current density $J_c$ (according to Bean's model $J_c \propto \Delta M$). [19] This is quite expected as various types of disorder act as the extra pinning centers and thus lead to higher $J_c$. Figure 4 show low field MH isotherms for all three compounds. The $H_{c1}$ values for $ZrTe_3$, $Cu_{0.05}ZrTe_3$, and $Ag_{0.05}ZrTe_3$, calculated from the point of deviation from the linear state in MH behavior are 6 Oe, 7.5 Oe and 8.5 Oe respectively. These values are similar to that for the $Cu_{0.05}ZrTe_3$ single crystals for H//ab. [4, 5] Though the field values taken from the point where magnetization deviates from the linearity do not represent the actual $H_{c1}$, but the field where magnetic field start penetrating into the sample. However for polycrystalline samples in powder form we can approximately take this value as $H_{c1}$. Similar to the Upper critical field $H_{c2}$, Lower critical field ($H_{c1}$) also shows linear dependence with temperature.

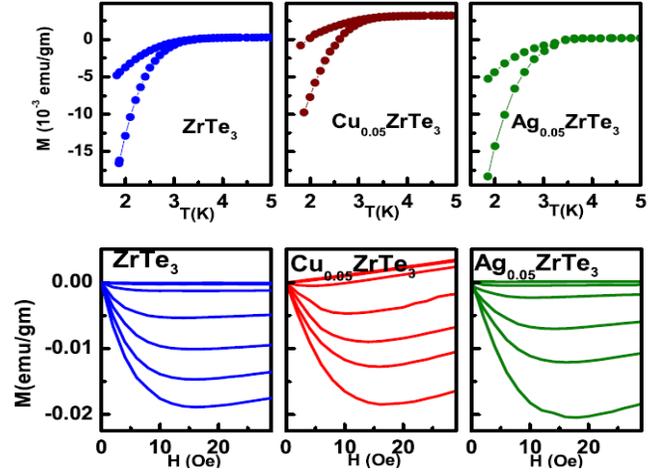

Fig. 4 (Color online) Temperature dependence of the DC magnetization (H=20 Oe) of $ZrTe_3$, $Cu_{0.05}ZrTe_3$, and $Ag_{0.05}ZrTe_3$ (upper panel). The magnetic field isotherms at temperatures T = 1.8 K, 2.0 K, 2.2 K, 2.5 K, 3.0 K, 3.5 K, and 4.0 K are shown in the lower panel for these compounds.

Specific heat (C) of the $ZrTe_3$, and $Ag_{0.05}ZrTe_3$ are shown in the figure 5. C vs T plot does not show any sharp jump. However similar to the HC reports for the single crystalline $ZrTe_3$, a weak jump in Heat capacity is seen in the C/T vs $T^2$ plot. We have shown C/T vs $T^2$ plot for $Ag_{0.05}ZrTe_3$ in the upper inset of the figure 5. The values of γ and β obtained from the low temperature intercept and slope of C/T vs $T^2$ curves in normal state respectively are γ = 0.6 mJ/mole-K$^2$, β= 2.95 mJ/mole-K$^4$ for $ZrTe_3$, γ = 3.35 mJ/mole-K$^2$, β = 0.97 mJ/mole-K$^4$ for Ag $ZrTe_3$. These values are slightly different from the reported values of γ = 3.20 mJ/mole-K$^2$ and β = 2.31 mJ/mole-K$^4$ for $ZrTe_3$ and γ = 2.64 mJ/mole-K$^2$ and β = 1.21 mJ/mole-K$^4$ for $Cu_{0.05}$ $ZrTe_3$ single crystals. [4, 20, 21] The evolution of phonon density of state, $(C-\gamma T)/T^3$, with the temperature shows a clear drop around SC transition (lower inset of figure 5). The DC magnetization at H = 0.5 T shows the paramagnetic behavior for ZrTe$_3$ and Ag$_{0.05}$ZrTe$_3$ (figure 6). Magnetic susceptibility of the metal depends on the electronic structure in the vicinity of the Fermi energy. For a metal with a high density of states D(E$_F$) at E$_F$, χ is dominated by the paramagnetic Pauli component χ$_P$ in the absence of the exchange enhancement and spin orbit coupling. χ$_P$(T) is proportional to the average density of states within ~k$_B$T of the Fermi level. We observe a drop in susceptibility at T$_{CDW}$ in ZrTe$_3$, Cu$_{0.05}$ZrTe$_3$, and Ag$_{0.05}$ZrTe$_3$, confirming of the change in the electronic structure near the Fermi surface. The onset of CDW leads to the opening of gap at Fermi surface and reduction of charge carriers, results in the drop of susceptibility. However the susceptibility anomaly is weaker for the Ag$_{0.05}$ZrTe$_3$, and Cu$_{0.05}$ZrTe$_3$ compared to ZrTe$_3$.

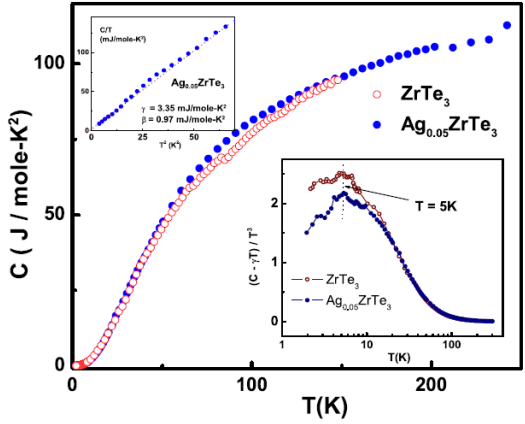

Fig. 5 (Color online) Temperature dependence of the Heat capacity of ZrTe$_3$, and Ag$_{0.05}$ZrTe$_3$. The top inset show the C/T vs T$^2$ for Ag$_{0.05}$ZrTe$_3$. The lower inset shows the temperature variation of phonon density of state indicating sharp variation at the SC transition.

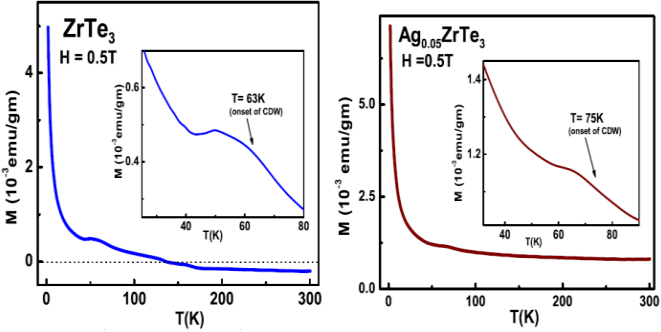

Fig. 6 (Color online) Temperature dependence of the magnetization for ZrTe$_3$ (left panel), and Ag$_{0.05}$ZrTe$_3$ (right panel) taken at 0.5 T magnetic field. The inset shows the enlarged portion of the curve showing the CDW anomaly, indicative of the depletion of conduction electrons.

The electrical resistivity (ρ) and its temperature derivative (dρ/dT) for the compounds are shown in figure 7, on an enlarged scale, to elucidate the CDW onset temperature. As seen from the figure, the onset of CDW ~ 80K (determined from the dρ/dT) remains same for all compositions and the full CDW sets in near 63K. The T$_{CDW}$ for polycrystalline ZrTe$_3$ is same as reported for the single crystal ZrTe$_3$. Though the intercalation of Cu or Ag does not affect the CDW state in our polycrystalline samples, the amplitude of CDW anomaly changes with intercalation.

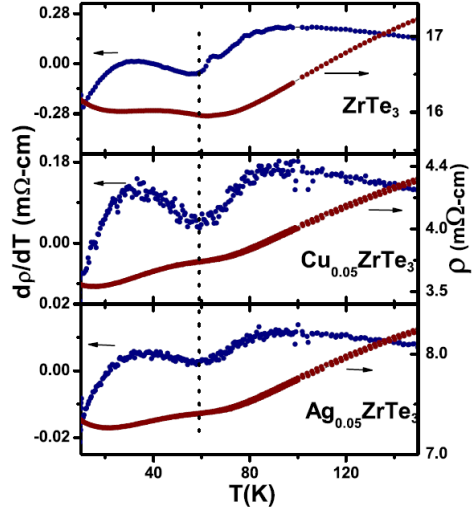

Fig. 7 (Color online) Temperature derivative of the resistivity (dρ/dT) for ZrTe$_3$, Cu$_{0.05}$ZrTe$_3$, and Ag$_{0.05}$ZrTe$_3$ is plotted along with the respective resistivity ρ(T) data to show the anomaly around the CDW transition.

In the figure 8, we have plotted the normalized electrical resistivity at room temperature for the ZrTe$_3$, Ag$_{0.05}$ZrTe$_3$, and Cu$_{0.05}$ZrTe$_3$ compounds to compare their normal state behavior. The resistivity increases down to 200 K upon cooling before decreasing again. The relative change in the slope of the resistivity is consistent with the room temperature resistivity values of these compounds. The CDW transition is clearly visible, and the resistivity rises again at lower temperature below CDW state, before collapsing at SC transition. We observed a bifurcation of resistivity data upon cooling and warming, above the CDW transition, the reason for which is not understandable to us, and need further study. Doping of the 5% Ag or Cu affects the CDW state slightly. Though there is no change in the T$_{CDW}$, the magnitude of CDW anomaly decreases upon Ag/Cu intercalation, which can be estimated from the size of resistivity anomaly (Δρ$_{CDW}$) at T$_{CDW}$. [13, 22] The change in resistivity at CDW, Δρ$_{CDW}$ is related to the reduction of the density of states at Fermi level (DOS(E$_F$)). The relative change in the resistivity due to CDW is given by 'α' = (R$_1$ − R$_2$)/R$_1$ = (σ$_1$ − σ$_2$)/σ$_1$, where R$_1$ and R$_2$ are the resistivities in the CDW state and the expected value in the absence of the CDW. [13, 22] The resistivity R$_1$ and R$_2$ are schematically shown in the inset of the figure 8. The conductivity of the metal is defined as σ = N$_0$e$^2$v$_f^2$τ, where N$_0$, v$_f$ and τ are the DOS at E$_F$, the Fermi velocity and the relaxation time of the conduction electrons, respectively. The formation of CDW state reduces N$_0$ but does not change v$_f$.

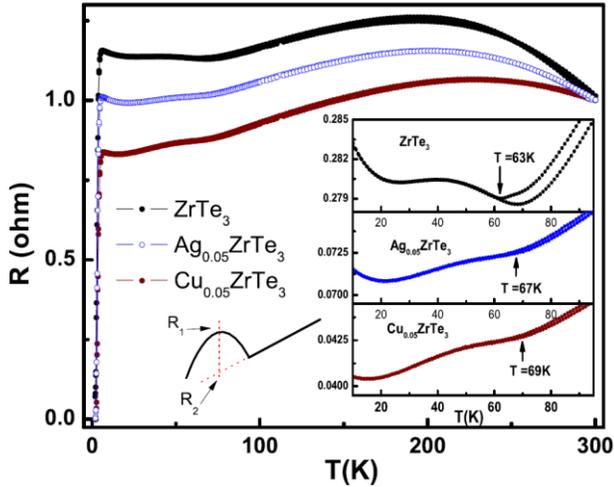

Fig 8 (Color online) Temperature dependence of the resistivity normalized with the room temperature value to compare the normal state nature of the ZrTe$_3$, Ag$_{0.05}$ZrTe$_3$, and Cu$_{0.05}$ZrTe$_3$. The presence of CDW anomaly is shown in the enlarged frame in the inset for the respective compounds.

Therefore, under the assumption that $\tau$ is not affected by the CDW formation $\alpha$ can be rewritten as '$\alpha$' = $(\sigma_1 - \sigma_2)/\sigma_1$ = $(N_0 - (N_0 - \Delta N))/N_0 = \Delta N/N_0$, where $N_0$ is the total density of states as $E_F$ in the normal state and $\Delta N$ denotes the reduction of $N_0$ due to the CDW formation. $\Delta N$ is proportional to the reduction of the Fermi surface due to the CDW formation, therefore '$\alpha$' can give the information of the size of the nested Fermi Surface. The values of '$\alpha$' for ZrTe$_3$, Ag$_{0.05}$ZrTe$_3$, and Cu$_{0.05}$ZrTe$_3$ are found to be 0.04, 0.027, and 0.025. It has been observed for the single crystal of ZrTe$_3$ that the pressure dependence of the superconducting transition temperature follows the pressure dependence of '$\alpha$'. [13] '$\alpha$' for the polycrystalline ZrTe$_3$, is same as for the crystals at 5 GPa pressure. [13] This is possible that the pressure due to strains in the polycrystalline compound is the reason for enhancement of $T_c$. In this context, it is to be mentioned that in NbSe$_3$ single crystals, superconductivity was never observed down to 50 mK and it is argued that a very high pressure is required for inducing superconductivity, the condition for which exist in the grain boundaries of polycrystalline sample. [14, 23, 24]

In conclusion, the polycrystalline samples of ZrTe$_3$ are found to superconduct at enhanced temperature of 5.2 K. The volume of superconducting fraction is less and we speculate that like the CDW state, which is observed in 'a' and 'c' directions only, superconductivity also sets in only along these directions in the remnant part of the Fermi surface, and therefore is not a bulk feature. Though it is filamentary in nature, It is akin to the effect of external pressure on $T_c$ in ZrTe$_3$ single crystal. The strains between the agglomerated small single crystallites in polycrystalline samples could be a possible reason for the higher Superconducting transition temperature $T_c$. Ironically we did not observe any effect of 5% intercalation of Cu and Ag, on either CDW or SC states. This is in contrast to the recent study in Cu intercalated ZrTe$_3$, where only intercalated phase is observed to show bulk superconductivity, whereas our polycrystalline sample of Cu$_{0.05}$ZrTe$_3$ and Ag$_{0.05}$ZrTe$_3$ does not show any improvement in the superconducting volume fraction. We have shown that Ag and Cu intercalation increases the electrical conductivity and affects the CDW anomaly, but has no bearing on the transition temperature of either CDW or SC phase.

We would like to acknowledge Manish Ghagh, and Prasad Mundye for providing help during the different stages of this work.
csyadav@tifr.res.in , paulose@tifr.res.in